\documentclass[openacc]{rstransa}

\newcommand\be{\begin{equation}}
\newcommand\ba{\begin{eqnarray}}
\newcommand\ee{\end{equation}}
\newcommand\ea{\end{eqnarray}}

\begin{document}

\title{Limitations of an Effective Field Theory Treatment
       of Early Universe Cosmology}

\author{
Robert Brandenberger$^{1}$}

\address{$^{1}$Physics Department, McGill University}

\subject{Cosmology, Early Universe, String Cosmology}

\keywords{Swampland, Trans-Planckian Censorship Conjecture, String Cosmology}

\corres{Robert Brandenberger\\
\email{rhb@physics.mcgill.ca}}

\begin{abstract}
 
Assuming that superstring theory is the fundamental
theory which unifies all forces of Nature at the quantum level,
I argue that there are key limitations on the applicability
of effective field theory techniques in describing early universe
cosmology.

\end{abstract}


\begin{fmtext}
\section{Introduction}

Most models of early universe cosmology are based on an effective field theory in which space-time is described in terms of General Relativity and matter is given by a set of fields. In this context, provided that the fields obey standard energy conditions, an initial cosmological singularity is unavoidable \cite{Hawking}. I will argue that, in this context, a breakdown of the effective field theory (EFT) desription of matter should be expected. One way to see this is to consider a field fluctuation with a fixed wavelength in comoving coordinates and trace it back in time. At some point, the wavelength becomes smaller than the horizon corresponding to the mass enclosed in the fluctuation region, and the fluctuation becomes a black hole, signalling the breakdown of the effective field theory description. Hence, in order to understand the evolution of the very early universe, we need to go beyond an effective field theory description. 

The correct understanding of the very early universe must be based on a theory of space, time and matter which does not suffer from the problems which effective field theories have at high energies. The best candidate we have is superstring theory. According to the perturbative formulation of string theory, the fundamental objects are not point particles but fundamental strings, and there is a fundamental length scale, the string length $l_s$. In any effective description of physics on length scales close to the string scale, the extended nature of the strings must lead to nonlocality.

In the following I will explore some avenues of connecting string theory with early universe cosmology.

\end{fmtext}


\maketitle

\section{Early Universe Models}

Before discussing consequences of superstring theory for cosmology it is important to remind the reader that the inflationary scenario \cite{Guth} is {\bf not} a necessary ingredient of early universe cosmology. There are, indeed, viable alternatives (see \cite{RHBalt} for a comparative review).

In this article, we work in the context of a homogeneous and isotropic background metric given by the line element
\be
ds^2 \, = \, dt^2 - a(t)^2 d{\bf{x}}^2 \, ,
\ee
where $t$ is physical time, ${\bf{x}}$ \footnote{Note that $k$ will stand for comoving wavenumber of a particular fluctuation mode.} are comoving spatial coordinates, and $a(t)$ is the cosmological scale factor in terms of which the Hubble expansion rate is given by $H(t) = {\dot{a}} / a$. The Hubble radius is defined as the inverse of the Hubble expansion rate. The Hubble radius plays an important role in the evolution of cosmological perturbations: perturbations on length scales larger than the Hubble radius are frozen out, whereas those on sub-Hubble scales can oscillate (see e.g. \cite{MFB, RHBfluctRev} for reviews). Hence, a local generation mechanism of fluctuations is only effective on sub-Hubble scales. The Hubble radius must be distinguished from the horizon (the forward light cone of some initial event) which yields the limit of causal contact. We use natural coordinates in which the speed of light $c$ and Planck's constant are set to $1$. Newton's gravitational constant is denoted by $G$, and the corresponding length scale is the Planck length $l_{pl}$. We will work in the context of a spatially flat universe. In this case, comoving coordinates can be normalized such that $a(t_0) = 1$, where $t_0$ is the present time.

Let us begin with the phenomenological requirements for a successful early universe theory.
Firstly, it must explain the near isotropy of the cosmic microwave background (CMB). Secondly, it must provide a causal mechanism for generating cosmological fluctuations with a nearly scale-invariant spectrum with a very small red tilt. In order to satisfy the first condition, there must be a phase in the early universe during which the horizon expands faster than the Hubble radius $H^{-1}(t)$ such that the horizon at the time $t_{rec}$ of recombination includes the entire spatial region which we see in the CMB. To obtain a causal structure formation scenario, comoving length scales which are probed today in CMB anisotropy experiments must originate inside the Hubble radius. In addition, the early universe model must yield a spectrum of fluctuations which is nearly space-invariant with a small red tilt (see e.g. \cite{RHBalt}).

The inflationary scenario \cite{Guth} has become the standard paradigm of early universe cosmology. It addresses some of the problems of Standard Big Bang cosmology and is the first model based on causal physics to yield an approximately scale-invariant spectrum of cosmological perturbations \cite{Mukh} and gravitational waves \cite{Starob}. Such a spectrum of fluctuations was known (based on earlier work of \cite{Zeld, Peebles}) to predict acoustic oscillations in the angular power spectrum of cosmic microwave background (CMB) anisotropies, and similar oscillations in the matter power spectrum.

Cosmological inflation \cite{Guth} indeed satisfies the abovementioned requirements: during the period of nearly exponential expansion of space the horizon expands exponentially while the Hubble radius is approximately constant (see Figure 1). Hence, provided that the period of inflation is sufficiently long, the horizon at   $t_{rec}$ becomes much larger than the radius of the spatial sphere which we see today with the CMB, and since the physical length associated with a fixed comoving scale also expands exponentially, it is possible that scales corresponding to the current Hubble radius start out smaller than the Hubble radius at the beginning of the period of inflation. In the case of an inflationary universe, it is reasonable to assume that the perturbations arise from quantum vacuum fluctuations. The resulting spectrum of curvature fluctuations \cite{Mukh, Press} and gravitational waves \cite{Starob} is then almost scale-invariant. The small red tilt of the spectrum stems from the fact that $H(t)$ decreases slowly in time. 
\begin{figure}[htbp]
\centering
\includegraphics[scale=0.53]{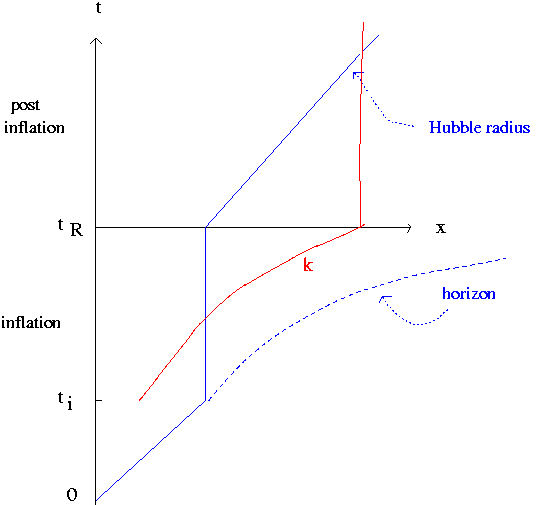}
\caption{Space-time sketch of inflationary cosmology. The vertical axis is time, the horizontal axis corresponds to physical distance. The inflationary phase of accelerated expansion lasts from time $t_i$ until time $t_R$. During this time interval and in the case of exponential expansion, the Hubble radius (the solid blue curve) is constant while the physical wavelength of a fixed comoving fluctuation mode (solid red curve marked by $k$) grows exponentially. Hence, modes can exit the Hubble radius. After inflation the universe evolves according to Big Bang cosmology, the Hubble radius grows linearly in time, and scales re-enter the Hubble radius.}
\end{figure}

The inflationary scenario is based on the assumption that there was a period in the early universe during which space expands (almost) exponentially. In general, cosmological inflation is studied at the level of an effective field theory based on General Relativity and scalar field matter. In this framework, inflation is know to be past-incomplete \cite{Borde}, suffering from an initial singularity problem. This is the first indication of the breakdown of the effective field theory description of early universe cosmology. As we will see below, the effective field theory description of inflation suffers from other conceptual challenges, and this motivates the search for alternative early universe scenarios.

There are alternative early universe scenarios which satisfy the abovementioned criteria (see e.g. \cite{RHBalt} for a more detailed discussion). Bouncing universe models can provide one such scenario. Here, time runs from $- \infty$ to $+ \infty$. The universe is initially in a contracting state. New physics yields a bounce at a time which can be scaled to be $t = 0$, and the bounce is followed by an expanding phase as described by Standard Big Bang oosmology. Since time is infinite in the past, the horizon is infinite. Thus, the first of the criteria is trivially met. As long as the contracting phase does not involve decelerated contraction, scales which we observe today, even though they are far beyond the Hubble radius at times near the bounce point, originated inside the Hubble radius at some early time (see Figure 2). The spectrum of fluctuations depends on the rate of contraction and on the assumption of their initial nature. If one posits that the fluctuations originate as quantum vacuum perturbations at early times, then it can be shown that a scale-invariant spectrum of curvature fluctuations at the end of the contracting phase is obtained provided that the equation of state of matter during the relevant stage of contraction is matter dominated \cite{Fabio}. In the case of ``Ekpyrotic'' (or ``ultra-slow'') contraction \cite{Ekp} vacuum perturbations yield scale-invariant fluctuations of certain metric fliuctuations, and that these can then yield scale-invariant curvature fluctuations after the bounce (see e.g. \cite{NewEkp, Ziwei} for different realizations).
\begin{figure}[htbp]
\centering
\includegraphics[scale=0.4]{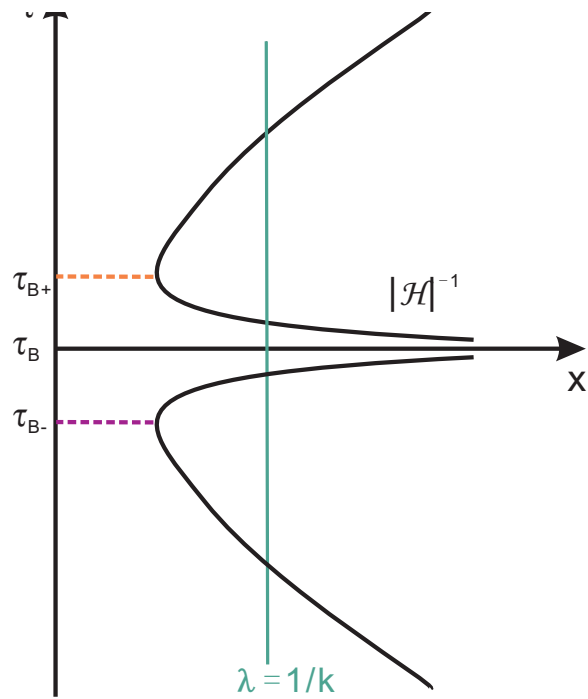}
\caption{Space-time sketch of a bouncing cosmology. The vertical axis is conformal time, the horizontal axis corresponds to comoving distance. The bounce happens at time $\tau_B$. Before that time, the universe is contracting, afterwards it is expanding. The comoving Hubble radius $|{\cal{H}}|^{-1}$ (shown as a solid black curve) decreases in the contracting phase, and thus scales (which have constant comoving wavelength) can exit the Hubble radius as they do in inflationary cosmology. However, the physical wavelength of the fluctuation mode is larger than the Planck length at all times, assuming that the energy scale of the bounce is smaller than the Planck scale.}
\end{figure}

Another class of alternatives to inflation are ``emergent'' cosmologies. Here, it is assumed that the expanding phase of Standard Big Bang cosmology begins after a phase transition from some early phase which has no standard effective field theory interpretation. String Gas Cosmology \cite{BV} is a toy model for such a scenario. Here, it is assumed that the early phase is a quasi-static phase of a gas of closed strings at a temperature close to the limiting temperature, the Hagedorn temperature \cite{Hagedorn}. The phase transition corresponds to the decay of winding strings into string loops, generating radiation. As in the case of a bouncing cosmology, the horizon is infinite, Since the Hubble radius is also infinite, all scales trivially emerge with sub-Hubble lengths (see Figure 3). It was shown that thermal fluctuations of a gas of closed strings yield a scale-invariant spectrum of cosmological fluctuations with a slight red tilt \cite{NBV}, and a scale-invariant spectrum of gravitational waves with a slight blue tilt \cite{BNPV}.
\begin{figure}[htbp]
\centering
\includegraphics[scale=0.45]{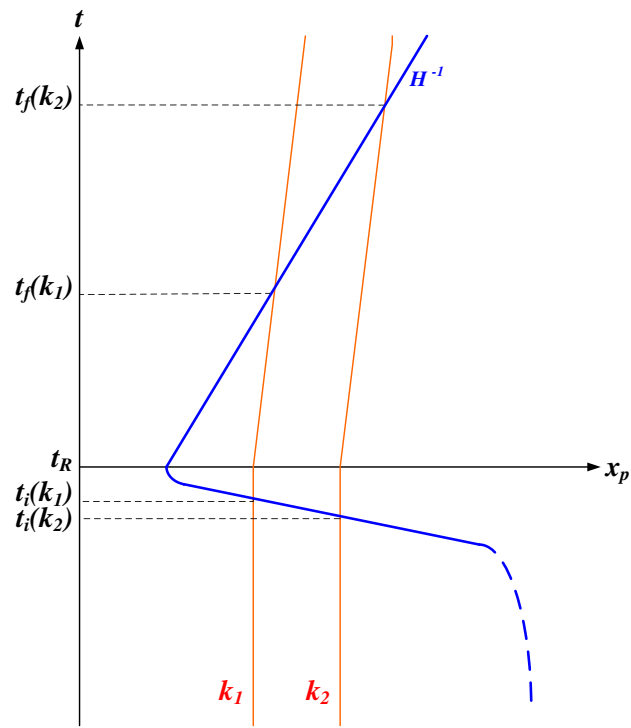}
\caption{Space-time sketch of an emergent cosmology. The vertical axis is time, the horizontal axis corresponds to physical distance. The transition between the early emergent phase (modelled here as being quasi-static) and the radiation phase of Standard Big Bang cosmology occurs at the time $t_R$. The Hubble radius is given by the solid bllue curve, while the two red curves labelled by $k_1$ and $k_2$ are the physical wavelengths of two comoving fluctuation modes. The physical wavelength of the fluctuation modes are larger than the Planck length at all times, assuming that the energy scale during the emergent phase (which sets the value of the Hubble radius at the time $t_R$) is smaller than the Planck scale.}
\end{figure}

\section{Swampland Criteria}

Which of the abovementioned early universe scenarios can emerge from superstring theory? This question has recently been the focus of a lot of attention, specifically concerning the possibility of obtaining cosmological inflation from string theory. Recent work indicates that there a number of key constraints on effective field theory models of inflation which can be consistent with superstring theory. These are called the {\it swampland constraints} (see \cite{Cumrun, Palti, Irene} for reviews).

The origin of the swampland constraints is the following: any scalar field in an effective field theory coming from superstring theory has a geometric origin in superstring theory. For example, the scalar field may be related to the radius of one of the extra spatial dimensions, or to the position of a D-brane. Hence, field ranges and potential energy functions are not freely tunable, but are determined by string theory.

The first swampland criterion is the {\it distance conjecture} \cite{OV} which states that the field range of a scalar field $\varphi$ emerging from string theory cannot vary by more than the Planck scale $m_{pl}$ (modulo a constant of the order of one). The reason for this condition is that  as $\varphi$ varies, the masses of some towers of string states decreases and eventually becomes so low that they have to be included in a new effective field theory. The distance conjecture constrains single field models of inflation with large field values, the models for which inflation is a local attractor in initial condition space (see e.g. \cite{RHBICrev} for a recent review of this issue).

A second criterion which is very constraining for inflation is the {\it de Sitter} conjecture, which states that potential energy functions $V(\varphi)$ for scalar fields must obey either \cite{dS1}
\be
\frac{V^{\prime}}{V} m_{pl} \, > \, c_1 \, 
\ee
(where a prime denotes the derivative with respect to $\varphi$) or \cite{Krishnan, dS2}
\be
\frac{V^{\prime \prime}}{V} m_{pl}^2 \, < - c_2 \, ,
\ee
where $c_1$ and $c_2$ are constants of order one. Thus, the potential needs to be either sufficiently steep or sufficiently tachyonic \footnote{These criteria are for positive potentials. See \cite{Heliudson} for a version of this criterion for negative potentials.}. Since canonical slow-roll models require a flat potential in order to obtain an equation of state which can lead to accelerated expansion, the de Sitter conjecture is a very serious constraint on inflationary models consistent with string theory, at least in the context in which these models can be described in terms of an effective field theory. Note that classes of models in which inflation can take place on steep potentials such as warm inflation \cite{warm} or models in which the scalar field driving inflation is coupled to other fields which slow down the motion of $\varphi$ (see e.g. \cite{gaugeflation}) can be consistent with the de Sitter constraint.

The de Sitter criterion also has implications for late time cosmology. As argued in \cite{SV}, it implies that Dark Energy cannot be a cosmological constant, and it also constrains quintessence models \cite{Lavinia}, although the constraints are still consistent with current observations.

The Ekpyrotic scenario \cite{Ekp}, one of the alternatives to inflation, is characterized by a period of super-slow contraction which is obtained by assuming that the dominant form of matter is a scalar field $\varphi$ with a steep negative exponential potential. Such exponential potentials are uniquitous in string theory constructions (see e.g. \cite{Baumann}). For example, the radius $r$ of one of the extra spatial dimensions of string theory yields a scalar field $\varphi$ via an exponential relation $\varphi / m_{pl} \sim {\rm{exp}}( r / l_s)$, and hence a power law potential energy function for the radion $r$ leads to an exponential potential for $\varphi$. The fact that a negative sign arises is related to the fact that the ground state in superstring theory is typically $AdS$. The steepness of the potential is consistent with the de Sitter criterion, as a generalization of the derivation of the de Sitter constraint to the case of negative potentials has shown \cite{Heliudson}.

\section{Trans-Planckian Censorship Conjecture}

In inflationary cosmology, the physical wavelength of cosmological fluctuations increases exponentially. Hence, if the phase of inflation lasts sufficiently long, then length scales corresponding to those which are currently measured in CMB anisotropy experiments and large-scale structure surveys originate at the beginning of the period of inflation with a physical length smaller than the Planck length. In this case, it is clear that an effective field theory description of the origin and evolution of these fluctuations must break down \cite{Jerome}. Motivated by these considerations, very recently another constraint on possible effective field theories of the early universe has been proposed, the {\it Trans-Planckian Censorship Conjecture} (TCC) \cite{Bedroya1}. It states that no effective field theory leading to this problem can be consistent with superstring theory. More specifically, the statement is that no scale which is initially (at some time $t_i$) trans-Planckian can ever become larger than the Hubble radius at a later time $t_R$. In terms of an equations, this means
\be \label{TCCeq}
\frac{a(t_R)}{a(t_i)} l_{pl} \, < H^{-1}(t_R) \, ,
\ee
for any initial time $t_i$ and final time $t_R$.

Another way to justify the TCC is \cite{RHB-TCC} is in analogy with Penrose's cosmic censorship hypothesis \cite{Penrose} which states that time-like singularities must be hidden from an external observer by an event horizon. The TCC can be viewed as a momentum space generalization of Penrose's hypothesis, the space of trans-Planckian modes playing the role of the black hole singularity and the Hubble horizon playing the role of the black hole event horizon.

For applications to inflation, one applies (\ref{TCCeq}) with $t_i$ being the beginning of inflation and $t_R$ the time of reheating. For inflation to yield a successful model of structure formation, the scale corresponding to the current Hubble radius must originate inside the Hubble radius at time $t_i$, i.e.
\be \label{successeq}
\frac{a(t_i)}{a(t_0)} H(t_0)^{-1} \, < \, H^{-1}(t_i) \, .
\ee
As discussed in \cite{Bedroya2}, the TCC criterion (\ref{TCCeq}) yields an upper bound on the duration of inflation, while (\ref{successeq}) yields a lower bound. These two bounds are only consistent if inflation takes place at a very low energy scale of $V^{1/4} < 10^9 {\rm{GeV}}$, assuming an almost constant value of $H$ during inflation and instantaneous reheating. This is a very severe constraint on canonical single scalar field models of inflation.  The bound can be relaxed by allowing a substantial time dependence of $H$ during inflation or by modifying the post-inflation cosmology \cite{TCCmod}. On the other hand, the bound is strengthened if there is a period of radiation domination before the onset of inflation \cite{Ed}.

As is obvious from Figs. 2 and 3, bouncing and emerging cosmologies are automatically consistent with the TCC as long as the energy scale at the bounce (or during the emergent phase) is smaller than the Planck scale. Of the early universe scenarios being considered at the moment, it is thus only the inflationary scenario which is in tension with the TCC.

Note that models in which the inflationary phase is a nonperturbative state built on top of a Minkowski background such as in the approaches of \cite{Dvali} and \cite{Keshav} may be safe from the TCC constraint, a point which has also been made in \cite{DvaliReply}.

Note that the TCC also has implications for late time cosmology. For example, is implies that Dark Energy cannot be due to a cosmological constant - it must be dynamical.

\section{Basics of String Cosmology}

In the two previous sections, I have argued that models of inflation based on an effective field theory treatment are in tension with superstring theory. While a phase of Ekpyrotic contraction can be obtained self-consistently from an effective field theory treatment, it is difficult to obtain a successful transition from contraction to expansion in the context of a stable EFT (but see \cite{Ijjas} for some recent attempts). In the recent realization of the Ekpyrotic scenario of \cite{Ziwei}, it is an S-brane, an object which does not have a smooth description in an EFT, which mediates the transition between contraction and expansion. Also, it is not possible to obtain an emergent phase such as that postulated in String Gas Cosmology, in the context of an EFT based on Einstein gravity coupled to matter which obeys of Null Energy Condition. All these evidences point to the need of going beyond an EFT analysis in order to understand the evolution of the early universe.

In the standard perturbative approach to superstring theory, the basic object of the theory are fundamental strings and not fundamental particles. From the point of view of a particle description, strings are nonlocal and hence the high energy description of string theory will not fit into the framework of a conventional EFT. A consistent model the very early universe can only be obtained if the model takes into account the crucial new degrees of freedom and new symmetries which are the basis of superstring theory.

{\it String Gas Cosmology} (SGC) \cite{BV} is an early attempt to set up a model of the early universe based on fundamental aspects of superstring theory. The new degrees of freedom, namely the string oscillatory and winding modes, which cannot be correctly included in a point particle based EFT, play a crucial role. For simplicity, consider space to be a nine-dimensional torus with radius $R$. Since the energy of string winding modes scales as $R$ while the energy of the momentum modes scales as $1/R$, and since the energy of the string oscillatory modes is independent of $R$, we have a key new symmetry of the theory, a symmetry under
\be \label{Tduality}
R \, \rightarrow \, \frac{1}{R} \,
\ee
(in string units). This symmetry is one manifestation of a larger symmetry group of string theory, the T-duality symmetry \cite{Tdual}. If we assume that matter is a gas of closed strings in thermal equilibrium (in analogy to how matter in Standard Big Bang cosmology is considered to be a gas of point particles in thermal equilibrium), then the symmetry (\ref{Tduality}) implies that the temperature $T$ obeys
\be
T(R) \, = \, T(\frac{1}{R}) \, .
\ee
There is no divergence of the temperature as $R \rightarrow 0$. In fact, the physics a $R \rightarrow 0$ is dual to the physics for $R \rightarrow \infty$ from the point of view of winding modes (which are the light degrees of freedom for small $R$) instead of from the point of view of the momentum modes (which are light for large $R$ but heavy for small $R$ \cite{BV}.

Due to the presence of the infinite tower of string oscillatory modes, there is a maximal temperature for a gas of closed strings in thermal equilibrium, the Hagedorn temperature $T_H$ \cite{Hagedorn}. As space contracts, then, once the temperature approaches this limiting value, instead of the energy of the initial string modes (the momentum modes) increasing, the energy flows into the oscillatory modes, maintaining approximately the same temperature. In the case of heterotic string theory, the form of $T(R)$ is sketched in Fig. 4. The width of the range over which the temperature is nearly constant depends on the total entropy of the system. This range of $R$ is what constitutes the {\it Hagedorn phase}.
\begin{figure}[htbp]
\centering
\includegraphics[scale=0.45]{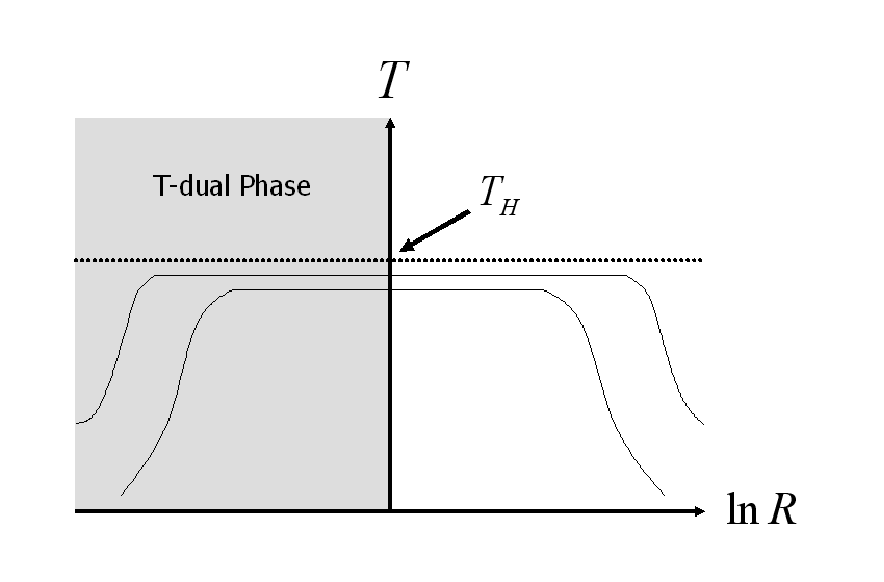}
\caption{The temperature $T$ as a function of the radius $R$ for a gas of closed heterotic strings in thermal equilibrium. $T_H$ is the Hagedorn temperature. The two curves correspond to different amounts of total entropy in the box: the higher the total entropy, the larger the range of values of $R$ over which $T$ remains close to the Hagedorn temperature.}
\end{figure}

During the Hagedorn phase, each dimension of the box is wrapped by winding strings. In the absence of a chemical potential, the net winding number about each cycle vanishes. But for winding modes to be able to annihilate, their world sheets need to interact. As argued in \cite{BV}, this only allows the annihilation to proceed in three of the nine spatial dimensions, the others being confined and stabilized at the string scale by the interaction of momentum and winding modes (see \cite{SGmoduli} for a detailed analysis). This nice feature of a dynamical origin of three large spatial dimensions relies crucially on going beyond an EFT description. Note that the decay of the string winding modes leads to string loops, i.e. to radiation. Thus, in SGC the transition from the early emergent phase to the radiation phase of standard cosmology is natural.
 
In the original model \cite{BV} it was assumed that the Hagedorn phase is quasi-static (including fixed dilaton). This leads to the space-time sketch of Fig. 3. In this case, it can be shown that thermal fluctuations of strings lead to a scale-invariant spectrum of curvature fluctuations with a slight red tilt \cite{NBV}, and to a scale-invariant spectrum of gravitational waves with a slight blue tilt \cite{BNPV}, a prediction with which this emergent scenario can be differentiated from that of the standard inflationary paradigm which predicts a slight red tilt (provided we work in the context of canonical inflation with matter obeying the usual energy conditions). More specifically, the dimensionless power spectrum ${\cal{P}}_{\Phi}$ of curvature fluctuations is
\be
{\cal{P}}_{\Phi}(k) \, \sim \, {\cal{A}} \epsilon^{-1}(k_p) \bigl( \frac{k}{k_p} \bigr)^{n_s - 1} \, ,
\ee
where $k$ is the wavenumber, $k_p$ stands for a pivot scale, and $n_s$ is the usual definition of the scalar spectral index. Similarly, the dimensionless power spectrum of gravitational waves is
\be
{\cal{P}}_{h}(k) \, \sim \, {\cal{A}} \epsilon(k_p) \bigl( \frac{k}{k_p} \bigr)^{n_t} \, ,
\ee
where $n_t$ is the usual definition of the tensor spectral index. In the case of standard inflation, we have $n_s < 1$ (i.e. a red scalar spectrum) and $n_t < 0$ (i.e. also a red spectrum), while the SGC scenario yields $n_s < 1$ and the consistency relation
\be
n_t \, = \, 1 - n_s \, > \, 0 \, ,
\ee
i.e. a blue tensor spectrum. The amplitude ${\cal{A}}$ is proportional to $(l_{pl} / l_s)^4$, and the quantity $\epsilon(k) = (1 - T(k) / T_H)$ measures how close to the Hagedorn temperature the temperature in the emergent phase is at the time when the mode with wavenumber $k$ crosses the Hubble radius towards the end of the emergent phase (see Fig. 3). It is the $k$ dependence of $T$ which leads to the tilts of the spectra for curvature fluctuations and gravitational waves. For detailed discussions on how these results are arrived at see \cite{SGCrevs}. 

The key assumptions which lead to the above results for the spectrum of fluctuations is that we have an emergent phase with thermal equilibrium. In this case, the perturbations will be of thermal origin (unlike in inflation where they are quantum vacuum fluctuations). The curvature perturbations are given by the energy density fluctuations of the thermal gas, and the gravitational waves are induced by the off-diagonal pressure perturbations. In connecting the matter fluctuations to the metric inhomogeneities we must make the assumption that the physics on infrared scales (the ones probed in current observations) reduces to the theory of linear perturbations in General Relativity, such that the formalism of \cite{MFB, RHBfluctRev} can be used. In order to obtain a scale-invariant spectrum for perturbations in three large spatial dimensions, it is important that the matter correlation functions have holographic scaling, i.e. they scale as the area of the box and not as the volume. The specifics of the emergent model are not important as long as these three criteria are met. We will in fact see in the following section that the criteria can be realizaed in a very different setting.

\section{Challenges and Approaches}

String Gas Cosmology provides a toy model of the very early universe which goes beyond what EFTs can yield. However, in \cite{BV} no dynamics for the emergnet phase, the Hagedorn phase, was provided. It is clearly of utmost importance to put the ideas of \cite{BV} on a firmer dynamical footing.

An early attempt at providing a dynamical background for string cosmology is {\it Pre-Big-Bang Cosmology} \cite{PBB} (see \cite{PBBrev} for an extensive review) which is based on the action of the massless bosonic modes of string theory. Setting the antisymmetric tensor field $B_{\mu \nu}$ to zero, this yields the equations of motion of dilaton gravity. The T-duality symmetry is reflected in a scale factor duality symmetry of the equations of motion. However, in this approach to date no phase has been discovered which could be a candidate for the Hagedorn phase expected from string theory. This is not surprising since this scenario is missing the key role which string oscillatory modes are expected to play.

A framework which has more recently been explored is {\it Double Field Theory} (DFT) (see \cite{DFT} for original articles and \cite{DFTrev} for a review). DFT is based on treating string momentum and winding modes on an equal footing, but within a field theoretical setting. This involves doubling the number of spatial dimensions, with coordinates $x^{i}$ \footnote{Here the Latin index $i$ runs over the usual 9 spatial dimensions, and Greek indices will run over the usual nine spatial dimensions and time.} representing space as seen from the point of view of the string momentum modes (i.e. the ``usual'' spatial dimensions), and dual spatial coordinates ${\tilde{x}}^i$ representing space as seen from the point of view of the string momentum modes. The usual metric $g_{\mu \nu}$ and  can be combined into a generalized metric whose indices run over the doubled space coordinates. For vanishing $B_{\mu \nu}$ (which is what is expected for homogeneous and isotropic cosmology), the generalized metric becomes block diagonal, with the block corresponding to the usual spatial dimensions being given by $g_{\mu \nu}$, and the block corresponding to the dual dimensions being given by its inverse, this being consistent with the expectation that distance seen from the point of view of the winding modes is the inverse of the distance seen from the point of view of the momentum modes. The classical action for the generalized metric is chosen such that one recovers the action for supergravity in the case when the fields only depend on the $x^i$ coordinates. 

The implications of DFT for cosmology have recently been explored in a series of works. At a classical level, geodesic completeness of the dynamics was argued for in \cite{DFT1}, and homogeneous and isotropic solutions of the DFT equations coupled to a string matter source were explored in \cite{DFT2}. The string matter source is meant to represent the effect of a gas of strings. As such, the matter source can be described in terms of a time-dependent equation of state $w(t)$, where $w$ is the ratio of pressure and energy density. To make contact with SGC, we assume that at early times $w(t) = - 1/d$ (the equation of state of string winding modes in $d$ spatial dimensions), and then at later times transits to $w(t) = 1/d$, the equation of state of radiation.

In \cite{HZ}, all possible worldsheet $\alpha^{\prime}$ (where $\alpha^{`}$ is the inverse of the string tension)  correction terms to the action were classified in the case of a homogeneous and isotropic vacuum background, and the analysis was generalized to include matter sources in \cite{BBF1}, and various types of solutions were explored. In \cite{BBF2}, the analysis was generalized to Bianchi space-times with separate scale factors for a three-dimensional external space and a six-dimensional internal space, and separate equations of state of the matter source in the external and internal directions. If we assume that at early times $t \ll 0$ the equation of state corresponds to winding modes about all spatial dimensions (i.e. $w_e = w_i = - 1/9$, and that at late times $t \gg 0$ the winding modes about the external dimensions annihilate into radiation, yielding $w_e = 1/3$, while equilibrium between momentum and winding modes about the internal dimensions yields $w_i = 0$, then solutions can be found for which the internal dimensions are static in the Einstein frame (except for a small decrease in the scale factor localized near $t = 0$, while the dynamics of the external dimensions is as postulated in SGC, namely constant scale factor in the early phase and radiation-dominated expansion in the late phase. The dynamics is nonsingular in both the String and Einstein frames. Note, however, that - unlike what is assumed in \cite{NBV}, the dilaton is not constant in the early phase. Since the formalism of \cite{HZ, BBF1} applies only in the case of homogeneous spaces, it is an open problem to compute the spectrum of cosmological perturbations which will emerge from the setup of \cite{BBF2}.

Another approach to an improved description of the emergent phase of a stringy early universe was put forwards in \cite{Vafa}. It is also based on the view \cite{BV} that string momentum modes and string winding modes must be treated on an equal level. The spatial dimensions we observe now are those associated with the momentum modes. As the radius of space (measured in terms of these modes) becomes smaller than the string length, then these modes become heavy while the winding mode sector becomes the one with the light degrees of freedom. It was argued in \cite{Vafa} that in the emergent phase, the description of the momentum mode sector should be in terms of a topological theory, and it was shown that with this assumption a scale-invariant spectrum of fluctuations emerges.

A very different approach to early universe string cosmology has  recently been suggested in \cite{Brahma}. The starting point of this analysis is the IKKT matrix model \cite{IKKT}, a proposed non-perturbative definition of superstring theory. The model is given by the action
\be \label{IKKTaction}
	S \, = \,  -\frac{1}{g^2} {\rm{Tr}} \bigl( \frac{1}{4} [A^a, A^b][A_a,A_b] + \frac{i}{2} \bar{\psi}_\alpha ({\cal{C}} \Gamma^a)_{\alpha\beta} [A_a,\psi_\beta] \bigr)\,
\ee
which involves 10 bosonic $N \times N$ Hermitean matrices $A^a$ ($a = 0, ... , 9$) and 16 fermionic matrices $\psi_{\alpha}$. $C$ is the charge conjugation matrix, and $\Gamma^a$ are the gamma-matrices for $D = 10$. The proposal of \cite{IKKT} is that in the limit $N \rightarrow \infty$ with $g^2 N$ fixed, the action (\ref{IKKTaction}) yields a non-perturbative definition of string theory. Note that the Latin indices are raised and lowered with the usual $\eta_{ab}$ matrix of special relativity.

In this approach, space, time and the usual matter fields are all emergent concepts. Since the matrices $A^a$ are  Hermitean, we can diagonalize one of them, and we choose $A^0$. We can also choose a basis in which the diagonal elements are in ascending magnitude. In the limit $N \rightarrow \infty$ this gives us continuous time $t$, where $t$ is obtained by averaging the diagonal elements of$A^0$ over a range $n$ of values (where $n \ll N$ but $n$ being a fixed fraction of $N$ as $N \rightarrow \infty$). The key observation of the work summarized in \cite{Nishimura} is that in the large $N$ limit, the matrices $A^i, i = 1, ...., 9$ become block diagonal, with elements outside of $n \times n$ blocks being greatly suppressed. For time $t$, we have spatial blocks ${\bar{A}}^i(t)$ which in the limit $N \rightarrow \infty$ become continuous space.

Given the thermal partition function for the matrix model, we can compute fluctuations of the energy-momentum tensor, which then induce late time curvature fluctuations and gravitational waves. The method is analogous to the one used in SGC \cite{NBV}: assuming that on infrared scales (the ones which we observe today) the equations of motion for fluctuations reduce to those of linearized General Relativity, the curvature perturbations and gravitational waves are given by the energy density fluctuations and the off-diagonal pressure perturbations, respectively, via the equations
\be \label{scalarfluct}
\langle\vert \Phi(k) \vert^2\rangle \, = \, 16 \pi^2 G^2 k^{-4} \langle \delta T^0_0(k) \delta T^0_0(k) \rangle \, ,
\ee
and 
\be \label{tensorfluct}
\langle\vert h(k) \vert^2\rangle \, = \, 16 \pi^2 G^2 k^{-4} \langle \delta T^i_j(k) \delta T^i_j(k) \rangle \,,\,\, i \neq j \, ,
\ee
where $k$ is the wavenumber of the fluctuation. Here, $\Phi$ is the the relativistic generalization of the Newtonian gravitational potential which describes the curvature perturbations, and $h$ is the amplitude of a gravitational wave mode. The results of a detailed calculation \cite{Brahma} demonstrate that the spectra of both curvature fluctuations and gravitational waves are scale-invariant, with the curvature spectrum obtaining a thermal component with a Poisson spectrum which dominates on very short wavelengths. In this calculation, the starting point is the BFSS matrix model \cite{BFSS} (for which the leading term in the high temperature limit is given by the IKKT model), and leading correction terms to the pure IKKT result play an important role.

The results obtained \cite{Brahma} in the context of the IKKT matrix model \cite{IKKT} are thus quite promising. In addition to obtaining emergent space and time, the model yields fluctuations which are similar in form to those obtained in SGC and hence in good agreement with current observations. Note that the Horizon Problem of Standard Cosmology is naturally solved in the context of this approach, as it is in the approach of \cite{Vafa} and in SGC. However, lots of work remains to be done to concretize the model, in particular to understand the transition between the emergent matrix model phase and late time cosmology.

\section{Conclusions and Discussion}

I have argued that, assuming that superstring theory yields the correct description of the four forces of nature at the quantum level, effective point particle field theory will break down in the early universe \footnote{See, however, \cite{Cliff} for an attempt to justify the use of effective field theory in cosmology.}. To correctly understand the primordial universe, an improved framework is needed. Given the challenges to inflationary cosmology discussed in Sections 3 and 4, one should seriously consider the possibility that the early universe did not undergo a period of inflationary expansion, and that the source of the structure in the universe is not quantum vacuum perturbations during an inflationary phase. In Sections 5 and 6 I have presented some attempts to construct an improved model of the early universe based on fundamental principles of string theory. These models do not fit into the framework of point particle effective field theories, but make use of the new degrees of freedom and new symmetries which string theory provides. The models presented here all are characterized by a stringy emergent phase, and both SGC and the matrix model cosmology of \cite{Brahma} have the property that thermal fluctuations in the emergent phase lead to a roughly scale-invariant spectrum of cosmological perturbations and gravitational waves, and thus fit the current data as well as inflationary models.

A lot of work remains to be done in order to put the models we have presented on a firmer footing. This area of mathematical cosmology hence has a bright future. 

\enlargethispage{20pt}




\competing{The author declares that he has no competing interests.}

\funding{The research was supported by a NSERC Discovery Grant and by the Canadian Research Chair program. Partial support from a Templeton Foundation subgrant is also acknowledged.}

\ack{I wish to thank my collaborators, in particular H. Bernardo, S. Brahma, R. Costa, G. Franzmann, S. Laliberte and A. Weltman for their contributions to our recent work. Most of the credit should go to them, but I take the blame for any misrepresentations. I also thank H. Bernardo and S. Brahma for comments on the manuscript.}



\end{document}